\documentclass[a4paper,11pt]{article}
\usepackage{pos}
\usepackage{gensymb}

\title{Radio-Morphing: a fast, efficient and accurate tool to compute the radio signals from air-showers}
 \ShortTitle{Radio-Morphing: a semi-analytic tool to compute the radio signals from air-showers}

\author*[a]{Simon Chiche}
\author[b,a,e]{Olivier Martineau-Huynh}
\author[a,f]{Kumiko Kotera}
\author[c, d, a]{Matias Tueros}
\author[f]{Krijn D. de Vries}

\affiliation[a]{Sorbonne Universit\'{e}, CNRS, UMR 7095, Institut d'Astrophysique de Paris, 98 bis bd Arago, 75014 Paris, France}

\affiliation[b]{Sorbonne Université, Université Paris Diderot, Sorbonne Paris Cité, CNRS,Laboratoire de Physique Nucléaire et de Hautes Energies (LPNHE), Paris, France}

\affiliation[c]{IFLP - CCT La Plata - CONICET, Casilla de Correo 727 (1900) La Plata, Argentina}

\affiliation[d]{Depto. de Fisica, Fac. de Cs. Ex., Universidad Nacional de La Plata,  Casilla de Coreo 67 (1900) La Plata, Argentina}

\affiliation[e]{National Astronomical Observatories, Chinese Academy of Sciences, Beijing 100012, China}

\affiliation[f]{Vrije Universiteit Brussel (VUB), Dienst ELEM, Pleinlaan 2, B-1050, Brussels, Belgium}

\emailAdd{simon.chiche@iap.fr}
\emailAdd{martineau@lpnhe.in2p3.fr}
\emailAdd{kotera@iap.fr}

\abstract{Radio detection of air-showers is a mature technique that has gained momentum over the past decades. With increasingly large-scale experiments, massive air-shower simulations are needed to evaluate the radio signal at each antenna position. Radio Morphing was developed for this purpose. It is a semi-analytical tool that enables a fast computation of the radio signal emitted by any air-shower at any location, from the simulation data of one single reference shower at given positions. It relies on simple electromagnetic scaling laws of the radio emission (i.e., electric field) at the antenna level and then an interpolation of the radio pulse at the desired positions.  We present here major improvements on the Radio Morphing method that have been implemented recently. The upgraded version is based on revised and refined scaling laws, derived from physical principles. It also includes shower-to-shower fluctuations and a new spatial interpolation technique, thanks to which an excellent signal timing accuracy of a fraction of nanosecond can be reached. This new implementation, provides simulated signals with relative differences on the peak-to-peak amplitude of ZHAireS simulations below 10\% (respectively 25\%) for 91\% (99\%) of antennas while the computation time was reduced by more than 2 orders of magnitude compared to standard simulations. This makes Radio Morphing an efficient tool that allows for a fast and accurate computation of air-shower radio signals. Further implementation of Askaryan emission or enabling to use an input value of the geomagnetic field should reduce relative differences with ZHAireS by few percents and make the method more universal.}

\FullConference{37$^{\rm{th}}$ International Cosmic Ray Conference (ICRC 2021)\\
		July 12th -- 23rd, 2021\\
		Online -- Berlin, Germany}

\begin{document}
\maketitle

\section{Introduction}

The preparation of up-coming large-scale radio experiments (e.g., GRAND~\cite{GRAND}, AugerPrime-Radio~\cite{Auger}, RNO~\cite{RNO}, IceCube-Gen2-Radio~\cite{IceCube}) requires to run a large number of air-shower simulations to evaluate their performances. While microscopic approaches such as Monte-Carlo simulations (ZHAireS~\cite{ZHAireS}, CoREAS~\cite{CoREAS}) provide an accurate way to model the radio-emission, these are highly time consuming with hours of computation needed to generate the radio signal of a single air-shower at the location of a hundred antennas. Macroscopic approaches are much faster but limited by our knowledge of the physical processes, particularly when looking for very inclined arrival directions. Radio Morphing, originally introduced in~\cite{radiomorphing}, is a semi-analytic tool that allows for a fast computation of any air-shower at any position from a small set of ZHAireS simulations used as references. The method is based on the principle of the universality of air-showers which assumes that the radio emission can fully be inferred by characterizing the shower at $X_{\rm max}$, supposed point like, such as variations in the radio emission between different showers can directly be modeled by accounting for the variations at $X_{\rm max}$. The electric field (i.e., the radio signal) of a microscopically simulated reference shower is first scaled using simple electromagnetic scaling laws. It can then be evaluated at any 3D antenna position thanks to dedicated interpolation methods. We present the major improvements that enable to overcome serious limitations of the initial version~\cite{radiomorphing}, in particular in terms of scaling accuracy, radio signal timing and shower-to-shower fluctuations. 

\section{Basic principles of the Radio Morphing}\label{section:rm_principles}

\begin{figure}[tb]
\centering 
\includegraphics[width=0.95\columnwidth]{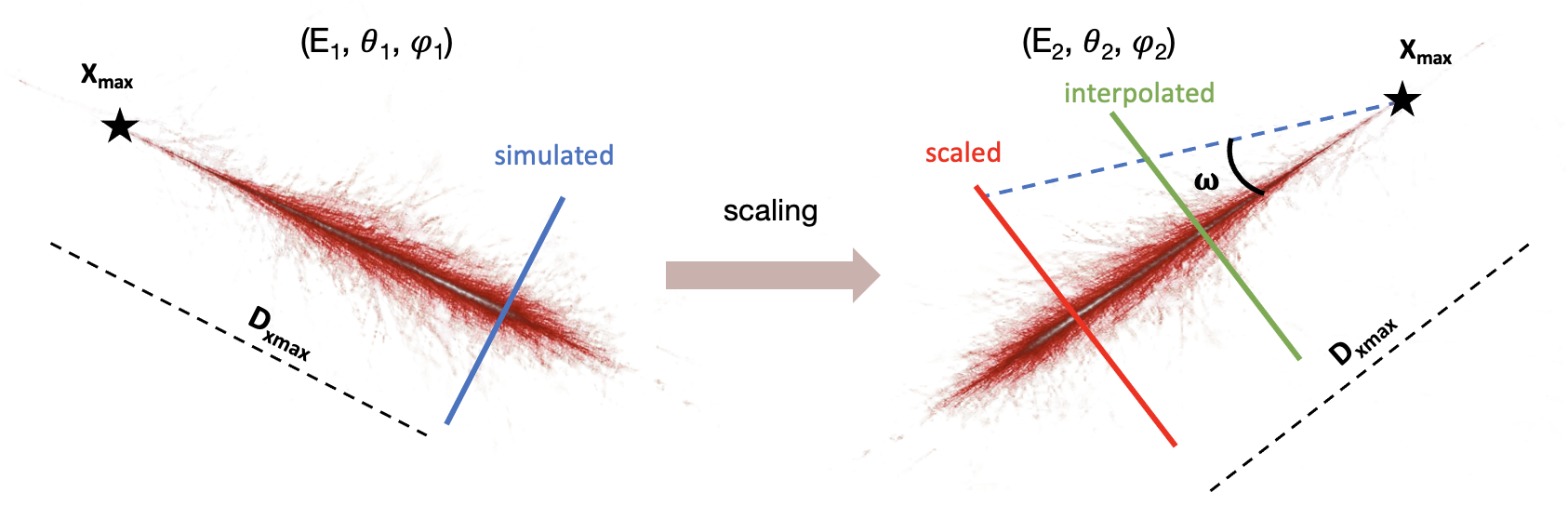}
\caption{Sketch of the Radio Morphing procedure. The electric field from a reference ZHAireS simulation with parameter ($E_{1}, \theta_{1}, \Phi_{1}, d_{\rm xmax}$) is scaled towards a shower with ($E_{2}, \theta_{2}, \Phi_{2}, d_{\rm xmax}$), the resulting signal is then interpolated to infer the radio-emission at any position.}\label{fig:sketch_rm}
\end{figure}

 The Radio Morphing procedure consists in two steps (see Fig.~\ref{fig:sketch_rm}): (1) Scaling: this step enables to infer the radio emission from any reference shower to any target shower with different parameters, but with same Xmax distance (i.e., distance between Xmax and the core of the shower in the shower plane). From a reference simulation run for a set of antennas located in a place perpendicular to the shower axis ("shower plane"), with given  parameters (primary nature, energy and arrival direction) we apply simple scaling laws to the layout and the electric field traces and to their footprint. (2) Interpolation: we use the resulting traces of antennas in the scaled plane to interpolate the electric field at any other position in space.  
 
 The scaling process relies on physical principles to account for the variation of the electric field with shower parameters such as the primary energy $\mathcal{E}$, azimuth $\Phi$, and zenith angle $\theta$. Following~\cite{radiomorphing}, we recall that the scaling of the electric field $E$ with primary energy is given by $E^{t} = k_{e}E^{r}$ with $k_{e} = \mathcal{E}_{t}/\mathcal{E}_{r}$, where indices $r$ and $t$ refer to the reference and the target shower respectively. A scaling is justified by the fact the number of particles in air-shower scales linearly with the primary energy at the first order. The scaling with the azimuth angle is done by correcting for variations of the geomagnetic angle $\alpha = (\widehat{\mathbf{u_{v}}, \mathbf{u_{B}}})$, where $\mathbf{u_{v}}$ and $\mathbf{u_{B}}$ are respectively the direction of the shower axis and of the local magnetic field. As the geomagnetic emission is due to the Lorentz force, its electric field amplitude depends linearly on $\sin{\alpha}$, hence our scaling corrects the $\mathbf{v \times B}$ component of the electric field by $E_{v \times B}^{t} = k_{\rm geo}E_{v \times B}^{r}$ with $k_{\rm geo} = \sin{\alpha^{t}}/\sin{\alpha^{r}}$. In practice, $E_{v \times B}$ contains also a minor charge excess or Askaryan component which does not scale with $\sin{\alpha}$. A more rigorous implementation taking this effect into account is under progress. 
 
\section{Scaling with the zenith angle}\label{section:zenith_scaling}

One of the main steps of Radio Morhping consists in modeling the effect of the zenith angle of the incoming air-shower direction on the electric field. We identified 3 mains contributions: (1) a change in the geomagnetic angle (described in the previous section) (2) a change in the refractive index at $X_{\rm max}$ and (3) a change in the density at $X_{\rm max}$. We detail below steps (2) and (3). 

The variation of the refractive index $n$ at $X_{\rm max}$ was already modeled in~\cite{radiomorphing}. Such variations affect the Cerenkov feature of the radio-emission. It can be taken into account by applying a stretching factor to the antennas position: $\Vec{x_{t}} =k_{c}\Vec{x_{r}}$ where $k_{c} = \theta_{c}^{t}/\theta_{c}^{r}$ is given by the ratio of the Cerenkov angle of the target and of the reference shower. Due to energy conservation, the electric field traces can be expressed: $\Vec{E^{t}}(\Vec{x}, t) = j_{c}\Vec{E^{t}}(\Vec{x}, r)$ with $j_{c}= 1/k_{c}$. 

\begin{figure*}[tb]
\includegraphics[width=0.49\linewidth]{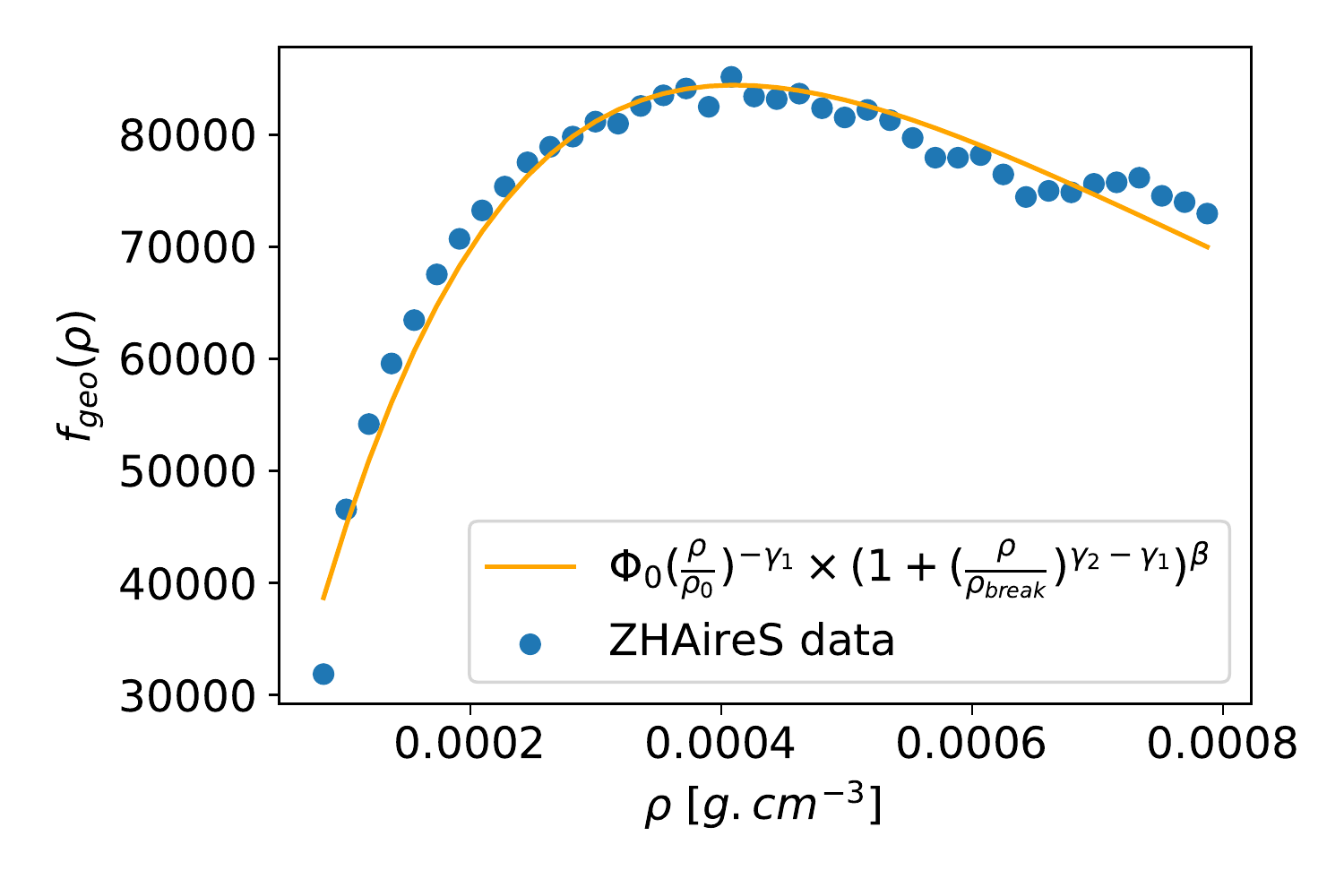}\hfill
\includegraphics[width=0.49\linewidth]{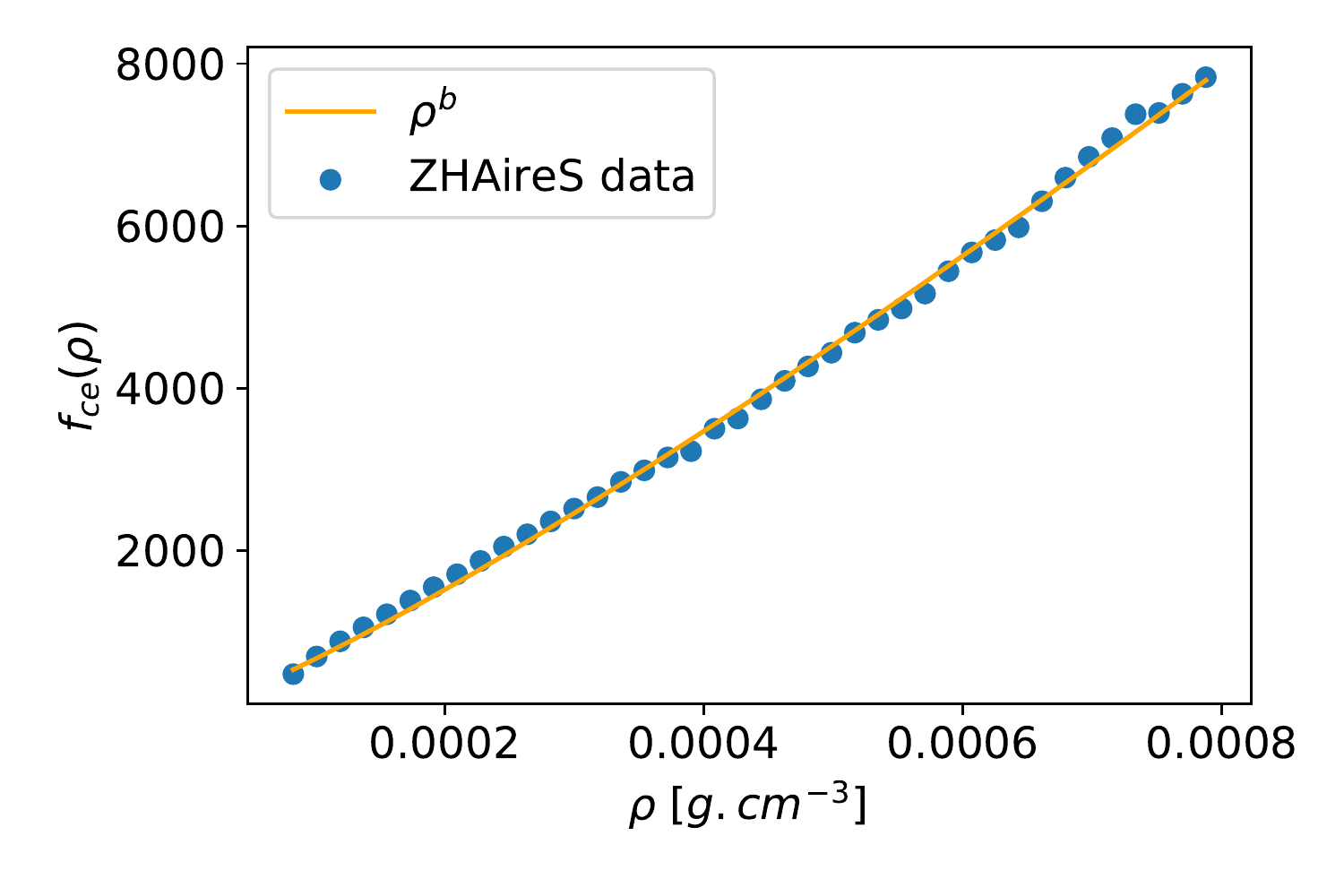}
\caption{({\it Left}) Fit of the geomagnetic electric field dependency with air-density with a broken a power law using $\phi_{0} = 1010$, $\gamma_{1} = -1.0047$, $\gamma_{2} = 0.2222$, $\rho_{\rm break} = 3.5\times 10^{-4}$, $\beta = 2.5$  and $\rho_{0} = 1.87\times 10^{-6}$. ({\it Right}) Fit of the charge excess electric field dependency with air-density with a simple power law using b = 1.195.}\label{fig:fit_ce_geo}
\end{figure*}

Variations of the zenith angle will also change the density at $X_{\rm max}$ as inclined showers develop in thinner atmosphere than vertical ones. Yet, only scarce literature can be found about the dependency of the geomagnetic and charge excess emissions with air-density, in particular for very inclined air-showers. We study this dependency using a set of $11\,000$ ZHAireS simulations of cosmic-rays on a star-shape layout with zenith angles between $45\degree$ and  $90\degree$, energies between $\rm 0.01\,EeV$ and $\rm 3.98\,EeV$ and 4 azimuth angles ($0\degree, 90\degree, 180\degree, 270\degree$). For each air-shower, we computed the geomagnetic and charge excess electric field amplitudes along the $\mathbf{v \times v \times B}$ baseline of antennas. Following Eq. 2.8 of~\cite{GlaserJCAP} and assuming a radial symmetry of the radio emission, we reconstructed the charge excess and geomagnetic radiated energies $E_{\rm rad,\, ce}$ and $E_{\rm rad,\, geo}$. Finally, these energies were then corrected from any dependency on the primary particle energy and on the geomagnetic angle via the following operations: $\tilde{E}_{\rm rad,\, geo} = E_{\rm rad,\, geo}/(\mathcal{E} \times \sin{\alpha})^{2}$ and $\tilde{E}_{\rm rad,\, ce} = E_{\rm rad,\, ce}/\mathcal{E}^{2}$. The residual dependency of $\tilde{E}_{\rm rad,\, geo}$ and $\tilde{E}_{\rm rad,\, ce}$ should then stem from density effects. Hence, assuming the atmospheric model of Linsley, we fit the geomagnetic and charge excess electric field amplitude dependency with air-density at $X_{\rm max}$, $f_{\rm geo}(\rho_{\rm xmax})$ and $f_{\rm ce}(\rho_{\rm xmax})$, which are illustrated in Fig.~\ref{fig:fit_ce_geo}. Note that the dependency between the density at $X_{\rm max}$ and the shower zenith angle can be easily computed numerically from a given injection height, $X_{\rm max}$ value and atmospheric model. In the initial version of the Radio Morphing~\cite{radiomorphing}, a scaling of the geomagnetic electric field amplitude in $1/\sqrt{\rho(X_{\rm max})}$ was assumed. We find that this scaling is only valid up to $\theta < 70\degree$. For the most inclined showers however (i.e., above $70\degree$), the amplitude of the geomagnetic electric field  decreases with decreasing density, an unexpected feature that remains to be fully understood. The physical origin of this trend, likely due to lower number of particles in thinner atmospheres and hence milder resulting currents, is being investigated.

The $\mathbf{v \times v \times B}$ component of the electric field corresponds to charge excess only. Hence it can be directly scaled with $f_{\rm ce}$. The  $\mathbf{v \times B}$ component however consists of charge excess and geomagnetic contributions. A rigorous scaling would require to decompose $E_{v \times B}$ into contributions from both emissions and then independently scale each component. Yet, as such a decomposition is not possible for antennas close to the $\mathbf{v \times B}$ baseline, we assume for simplicity that $E_{v \times B}$ is made of geomagnetic emission only, a reasonable hypothesis for showers in the range of zenith angles that we consider~\cite{}. As a consequence, one can scale the $\mathbf{v \times B}$ component directly with $f_{\rm geo}$. Finally the scaling of the electric field with air-density can be expressed
\begin{equation}
    E_{v \times B}^{t} = \frac{f_{\rm geo}(X_{\rm max^{t}})}{f_{\rm geo}(X_{\rm max^{r}})}E_{v \times B}^{r} \qquad {\rm and} \qquad E_{v \times v \times B}^{t} =  \frac{f_{\rm ce}(X_{\rm max^{t}})}{f_{\rm ce}(X_{\rm max^{r}})}E_{v \times v \times B}^{r} \ .
\end{equation}

For more accuracy in the scaling procedure, a sample of reference showers can be used (e.g., 5 reference showers with zenith angles [67.8, 74.8, 81.3, 83.9, 86.5]), instead of a single one. A library with a denser sampling for inclined air-showers is favored, as small variations in zenith translate to large differences in terms of propagation. Using such a reference library, we were able to extend the validity range of the Radio Morphing from $\theta = 85\degree$ to $\theta = 90\degree$ in zenith angle (initial version) to $\theta = 60\degree$ to $\theta = 90\degree$ for this new version. 

The scaling procedure was tested by computing the integral of the radio signal of Radio Morphed simulations and comparing them with the integral of the analogous ZHAireS simulations. The mean relative differences of the electric fields are of $\lesssim 10\%$. 

 Once the scaling process is completed, antennas of the scaled plane in Figure~\ref{fig:sketch_rm} are used to interpolate the radio emission at any position in space. We refer as 2D interpolation an interpolation performed inside of the scaled plane and 3D an interpolation at a position outside of that plane, both are presented in Section~\ref{section:interpolation}.

\section{Shower-to-shower fluctuations}\label{section:shower_fluct}

Due to the probabilistic nature of interactions occurring in air-shower, the radio-emissions produced by primary particles with identical nature, energy and arrival direction present some variations between different showers that we refer as shower-to-shower fluctuations. These fluctuations are intrinsically modeled by Monte-Carlo simulations for which hadronic interactions are evaluated in a probabilistic way. 
Radio Morphing uses only a small sample of reference showers and hadronic interactions are not recomputed when generating new showers. This allows to considerably reduce the computation time but requires to further implement shower-to-shower fluctuations that are consistent with the observed data. In this section, we present our implementation that relies on the modeling of: (1) $X_{\rm max}$ fluctuations and (2) fluctuation of the number of particles in the shower.

\subsection{$X_{\rm max}$ fluctuations}

When generating a new shower with the Radio Morphing, from an input primary, zenith angle, injection height and assuming Linsley's atmospheric model, the grammage is numerically integrated until the $X_{\rm max}$ value of the primary is reached, which we use as our emission point. Hence, using our set of $\sim$ $11\,000$ ZHAireS simulations, we parametrized the mean $X_{\rm max}$ value for proton and iron induced air-showers using a function of the form $\langle X_{\rm max} \rangle = a\log{E[\rm TeV]} +c$. , with $a =57.4 $, $c =421.9 $ for protons and $a = 65.2$, $c = 270.6$ for iron nuclei. This parametrization is chosen instead of an empirical one to better approach the outputs of numerical simulations. Additionally, in order to model shower to shower fluctuations of $X_{\rm max}$, we also parametrized the $X_{\rm max}$ root mean square $\sigma_{\rm xmax}$ from ZHAireS simulations as $\sigma_{\rm xmax} = a + b/E^c$, with $a =66.5$, $b =2.84$, $c =0.48$ for protons and $a =20.9$, $b =3.67$, $c =0.21$ for iron  nuclei.  In Radio Morphing, shower-to-shower fluctuations are modeled by generating $X_{\rm max}$ values for proton and iron nuclei using a random Gaussian distribution with $\langle X_{\rm max} \rangle$ and $\sigma_{\rm xmax}$ as input parameters.

\subsection{Fluctuations of the number of particles}

\begin{figure*}[tb]
\includegraphics[width=0.49\linewidth]{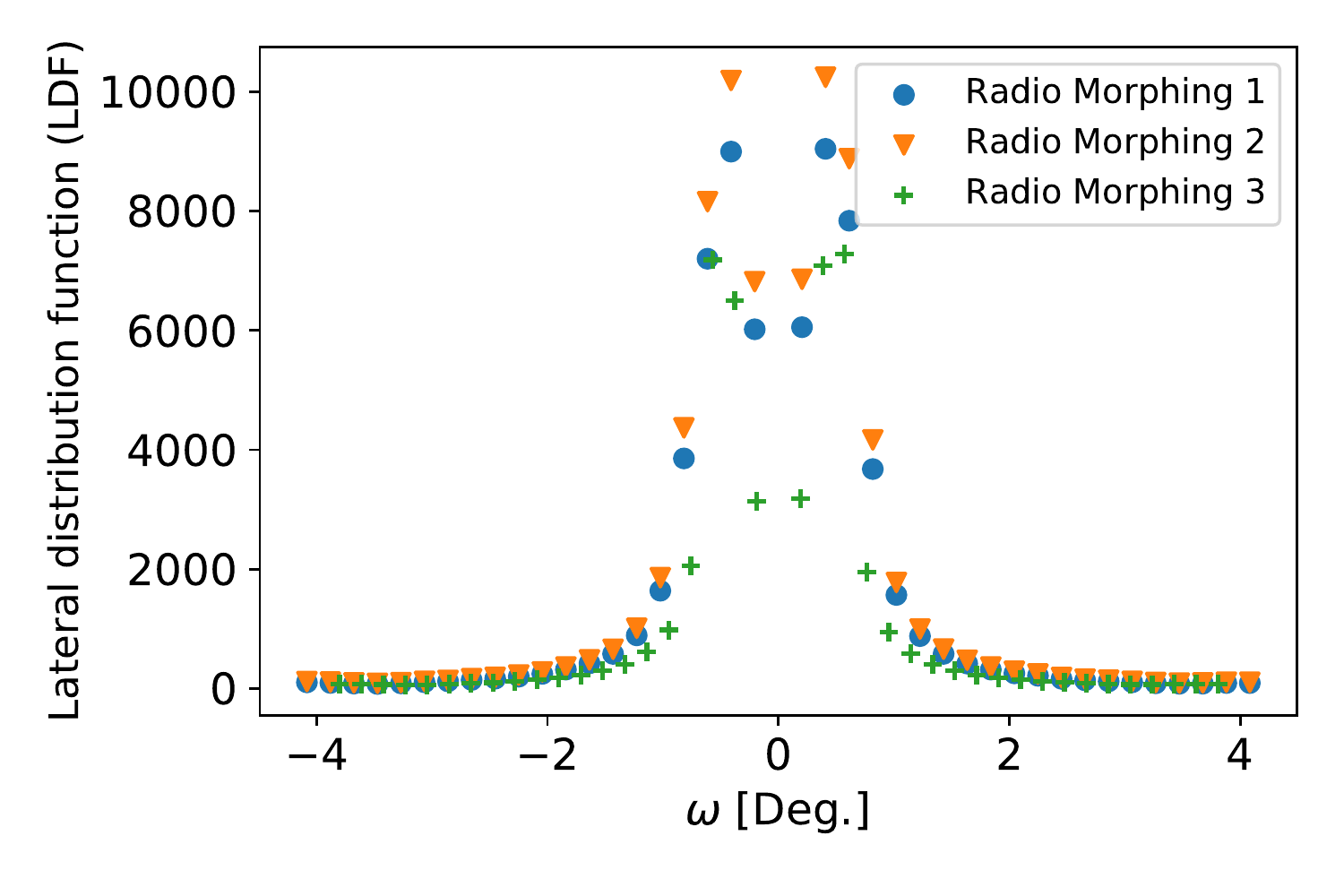}\hfill
\includegraphics[width=0.49\linewidth]{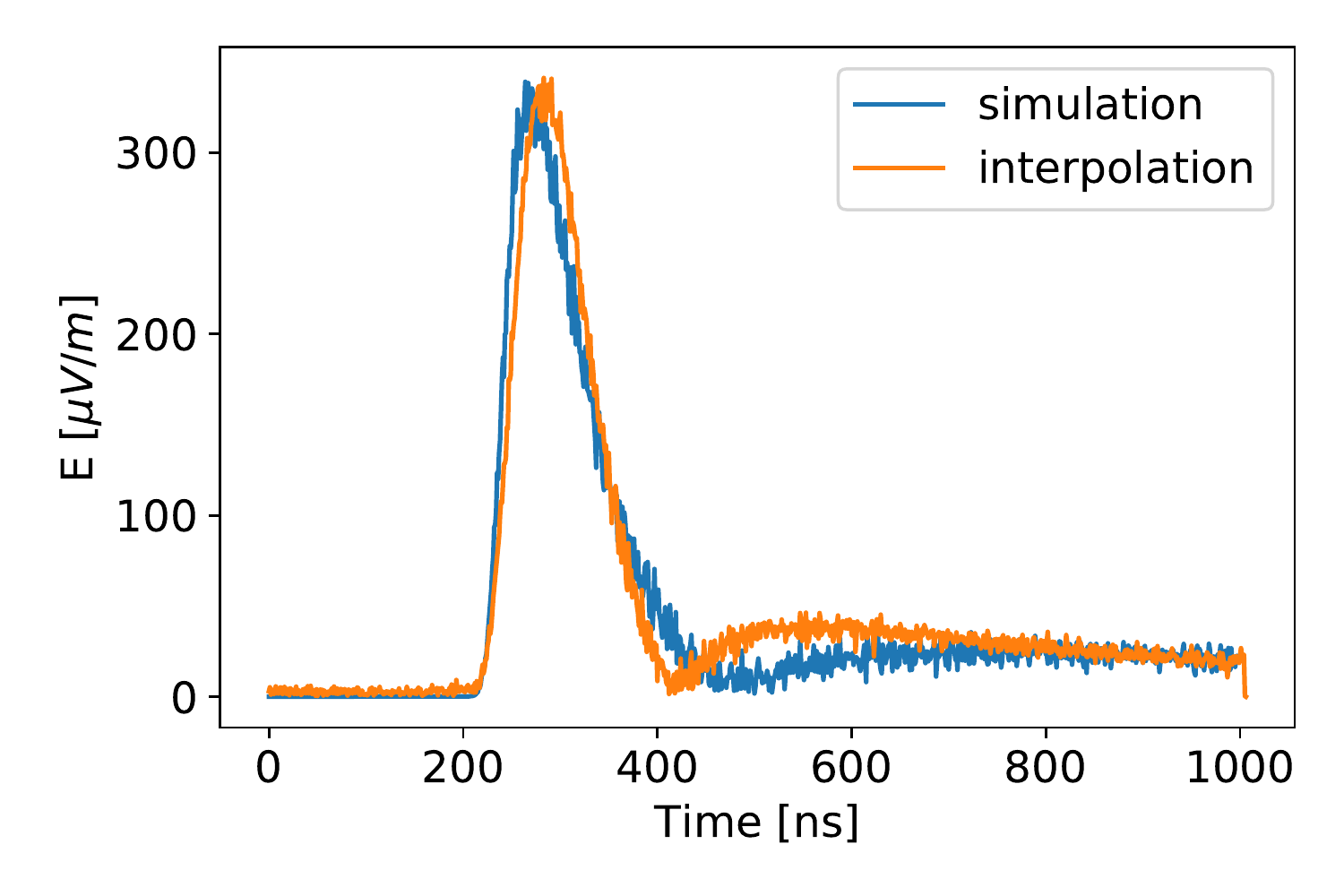}
\caption{({\it Left}) Lateral distribution function (LDF) of the electric field amplitude, along the $\mathbf{v \times B}$ baseline of antennas for 3 Radio Morphing outputs of a single reference shower with $\mathcal{E} = 3.98\,EeV$, $\theta = 85\degree$, $\Phi = 270\degree$ (towards East) with shower to shower fluctuations on. The horizontal axis $\omega$, corresponds to the angle between the shower axis and the direction that goes from $X_{\rm max}$ to a given antenna.
({\it Right}) Example of an interpolated trace after a scaling procedure along the East-West channel of an antenna for a shower with $\mathcal{E} = 3.98\,EeV$, $\theta = 85\degree$, $\Phi = 270\degree$. }\label{fig:fluctuation_interpolation}
\end{figure*}

We studied the fluctuation of the number of particles in air-showers using ZhaireS simulations: we computed the number of positrons and electrons, $N_{e^{+}e^{-}}$ that cross a plane perpendicular to the shower axis at $X_{\rm max}$ for showers with various arrival directions but fixed energy. We find fluctuations of roughly 10\% independently of the zenith angle or the primary particle energy. $N_{e^{+}e^{-}}$ fluctuations can directly be translated with Radio Morphing fluctuations of the primary particle energy as as it scales linearly with $N_{e^{+}e^{-}}$. Hence shower to shower fluctuations are modeled by randomly drawing a shower energy in a Gaussian distribution of 10\% sigma value around the target value $\mathcal{E}$. 
Shower to shower fluctuations are illustrated in the left panel of Fig.~\ref{fig:fluctuation_interpolation}. It can be seen that the fluctuations do not impact the shape of the LDF but still impact the amplitude of the electric field mostly for antennas with the highest signal.

\section{Interpolation}\label{section:interpolation}

Once the scaling procedure is completed, the next step of the Radio Morphing consists in using the radio-emission from the scaled antennas to then interpolate the electric field amplitude at any given position. The interpolation in itself can be divided into two main steps, first a 2D interpolation to infer the signal at any position included in the plane of the scaled antennas and then a 3D extrapolation to infer the radio-emission at any position outside of that plane. We present below these 2 steps.

\subsection{2D interpolation}

We use a 2D interpolation use the 2D interpolation method presented in~\cite{interpolation}. This method is performed in Fourier space where the electric field traces of antennas used for the interpolation are first decomposed into a phase and an amplitude, which are then linearly interpolated at the desired position. Finally, an inverse Fourier transform is performed to infer the desired traces. When considering antennas outside of the Cerenkov cone, this method provides relative differences with ZHAireS simulations of only a few percent on both the peak-to-peak amplitude and the integrated pulse. The interpolation method also provides a good agreement in timing with simulations, with an accuracy of less than 1 nanosecond. This is one of the important improvements of this version, compared to the initial Radio Morphing implementation~\cite{radiomorphing} which was not providing a correct timing information.

\subsection{3D: correcting for propagation effects}

The 3D interpolation is mainly an extension of the 2D, extrapolating beyond the shower plane by correcting for propagation effects. In Fig.~\ref{fig:sketch_rm}, $\omega = (\widehat{\mathbf{u_{v}}, \mathbf{u_{\rm antenna}}})$ represent the angle between the shower direction $\mathbf{u_{v}}$ and the direction that goes from $X_{\rm max}$ to a given antenna $\mathbf{u_{\rm antenna}}$. To infer the electric field of an antenna in the "interpolated" plane at a given $\omega$, we first perform a 2D interpolation to infer the electric field of the antenna at the same $\omega$ in the "scaled" plane, and then correct from propagation effects between the 2 planes. The correction of the propagation effects consists in first accounting for the dilution of the radio signal between the 2 planes. This can simply be done by applying a dilution factor $E_{\rm interpolated} = k_{d}E_{\rm scaled}$, where $k_{d} = D_{\rm scaled}/D_{\rm interpolated}$ corresponds to the ratio between the distance from $X_{\rm max}$ to the interpolated plane and the distance from $X_{\rm max}$ to the scaled plane. A second correction of the propagation consists in accounting for the difference in the integrated refractive index $\bar{n}$ between both planes. This is done by applying a stretching factor to the antennas positions and the traces $k_{\rm stretch} = \theta_{c}^{\rm interpolated}/\theta_{c}^{\rm scaled} = (\arccos{1/\bar{n}_{\rm interpolated}})/(\arccos{1/\bar{n}_{\rm scaled}})$, similarly to what was done for the scaling with the zenith angle. 

In Fig.~\ref{fig:fluctuation_interpolation}, the comparison between a 3D interpolated signal for the East-West channel of a given antenna after scaling, and a ZHAireS simulation is presented. We found a good agreement between both signals with a relative difference of 0.4\% on the peak-to-peak amplitudes.

\section{Results}\label{section:results}

\begin{figure*}[tb]
\includegraphics[width=0.49\linewidth]{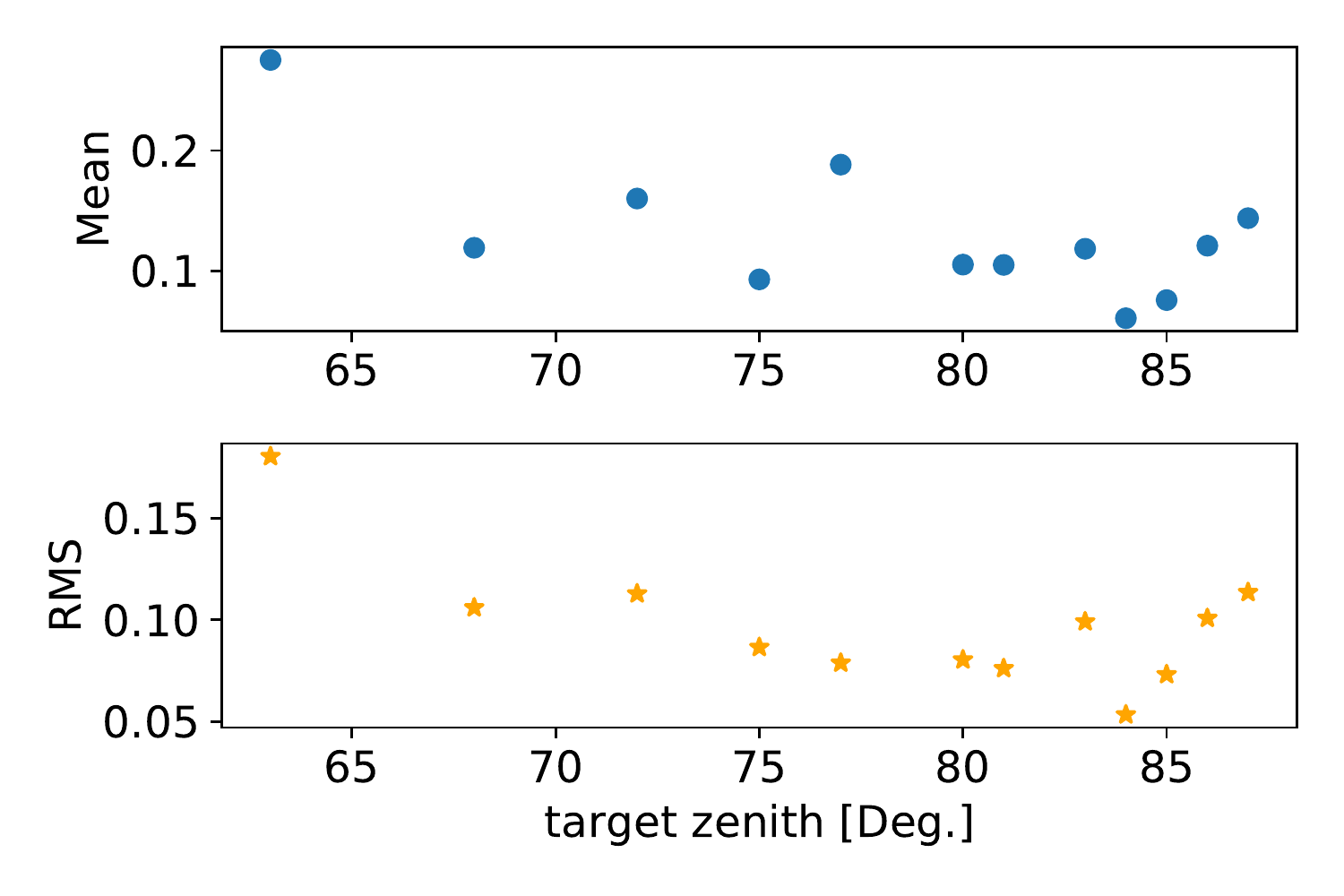}\hfill
\includegraphics[width=0.49\linewidth]{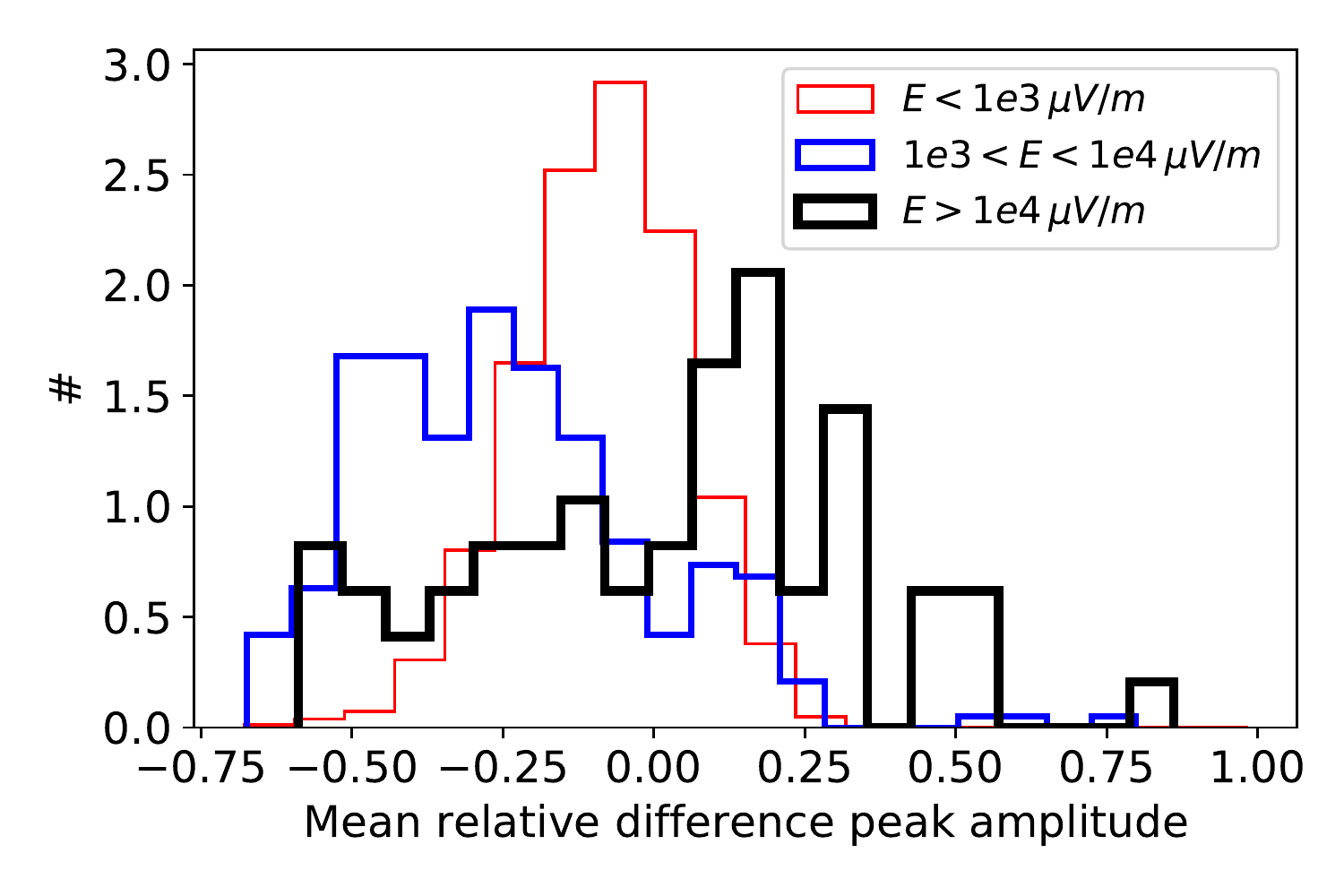}
\caption{({\it Left})  Relative differences on the peak-to-peak amplitude of Radio-Morphed antennas and corresponding ZHAireS simulations as a function of the target zenith angle for a set of $1200$ showers with 16 interpolated antennas each. ({\it Right}) Normalized histogram of the distribution of the error on the peak amplitude for 3 different ranges of the total electric field amplitude $E$.}\label{fig:result_radiomorphing}
\end{figure*}

The results of the full Radio Morphing method are presented in Fig.~\ref{fig:result_radiomorphing}. We considered a set of $\sim 1200$ cosmic-ray simulations with energies between $\rm 0.1\, EeV$ and $\rm 3.98\, EeV$,  zenith angles between $60\degree$ and $90\degree$ and 4 azimuth angles ($0\degree, 90\degree, 180\degree, 270\degree$). We consider planes of antennas perpendicular to the shower axis, the distance to $X_{\rm max}$ was chosen to have planes at $z\sim 1000\,m$ above see level. For each simulation, the electric field amplitude of 16 antennas was computed with Radio Morphing and then compared to the corresponding ZHAireS simulations.  On the left plot, we present the relative differences on the peak-to-peak amplitude of the total electric field between ZHAireS and Radio Morphing as a function of the zenith angle of the target showers. The blue points correspond to the mean values and the orange stars to the RMS. We obtain an accuracy between 5 to 20\% and a precision between 5 to 15\% for most data bins, independently of the zenith angle. In the right-hand plot of Fig.~\ref{fig:result_radiomorphing}, we show the normalized distributions of the mean errors for 3 different ranges of the total electric field amplitude. It appears that the largest errors come from This corresponds to antennas located on the Cerenkov cone for which the sharp amplitude variations limit the accuracy of the interpolation. We note however that the results presented here are full band, hence applying a frequency filter will reduce the Cerenkov features and improve the results. Also, the Radio Morphing method is sensitive to the reference library of showers that are also subject to shower to shower fluctuations. This limits the maximal precision that can be reached even with precise scaling procedure and interpolation. Nevertheless, the performances of this method are excellent, 91.5\% the Radio-Morphed signals at the antenna level have a relative error on the peak amplitude below 10\% and 99.1\% below 25\%.

\section{Conclusion}

Radio Morphing is an innovative tool that allows for a fast computation of any air-shower at any given position from a small set of ZHAireS reference simulations. It relies on two main steps: a scaling procedure based on simple physical principles to evaluate the impact of the parameters of the primary particle (nature, energy, arrival direction) on the amplitude of the electric field, and an interpolation which uses the signal of the scaled antennas to infer the radio-emission at any other position. We presented here novel implementations compared to the initial method presented in~\cite{radiomorphing}, to approach the experimental data. Specifically, the possibility to enable shower-to-shower fluctuations was added by parametrizing ZHAireS simulations. The scaling with the zenith angle of the shower arrival direction was also refined based on fits of the charge excess and geomagnetic emission dependencies with air density, extending the validity range of Radio Morphing to zenith angles of $60\degree$ to $90\degree$. Finally, a new interpolation method was implemented leading to relative differences with ZHAireS simulations of few \% on the amplitude, and timing accuracies of a fraction of nanosecond.

The full method (scaling + interpolation) provides results comparable to ZHAireS simulations, we find relative differences on the peak amplitude below 10\% (respectively 25\%) for 91.5\% (99.1\%) of antennas. In parallel, the computation time was reduced by several orders of magnitude compared to usual Monte-Carlo simulations. Further refinements such as correcting for the second order effects in the scaling with the geomagnetic angle, or enabling to use an input value for the geomagnetic field should provide an even more accurate and universal method.

%
%
%

\end{document}